\documentclass[twocolumn,floatfix, showpacs]{revtex4}

\usepackage[final]{epsfig}
\usepackage{amsmath}
\usepackage{parskip}
 
\begin{document}
 
\title{Parton shower evolution in a 3-d hydrodynamical medium}
 
\author{Thorsten Renk}
\email{trenk@phys.jyu.fi}
\affiliation{Department of Physics, P.O. Box 35 FI-40014 University of Jyv\"askyl\"a, Finland}
\affiliation{Helsinki Institute of Physics, P.O. Box 64 FI-00014, University of Helsinki, Finland}
 

\begin{abstract}
We present a Monte Carlo simulation of the perturbative Quantum Chromodynamics (pQCD) shower developing after a hard process embedded in a heavy-ion collision. The main assumption is that the cascade of branching partons traverses a medium which (consistent with standard radiative energy loss pictures) is characterized by a local transport coefficient $\hat{q}$ which measures the virtuality per unit length transferred to a parton which propagates in this medium. This increase in parton virtuality alters the development of the shower and in essence leads to extra induced radiation and hence a softening of the momentum distribution in the shower. After hadronization, this leads to the concept of a medium-modified fragmentation function. On the level of observables, this is manifest as the suppression of high transverse momentum ($P_T$) hadron spectra. We simulate the soft medium created in heavy-ion collisions by a 3-d hydrodynamical evolution and average the medium-modified fragmentation function over this evolution in order to compare with data on single inclusive hadron suppression and extract the $\hat{q}$ which characterizes the medium. Finally, we discuss possible uncertainties of the model formulation and argue that the data in a soft momentum show evidence of qualitatively different physics which presumably cannot be described by a medium-modified parton shower.
 
\end{abstract}
 
\maketitle

\section{Introduction}

Jet quenching, i.e.\ the energy loss of hard partons created in the first moments of a heavy ion collision due to interactions with the surrounding soft medium  has long been regarded a promising tool to study properties of the soft medium \cite{Jet1,Jet2,Jet3,Jet4,Jet5,Jet6}. The basic idea is to study the changes induced by the medium to a hard process which is well-known from p-p collisions. A number of observables is available for this purpose, among them suppression in single inclusive hard hadron spectra $R_{AA}$ \cite{PHENIX_R_AA}, the suppression of back-to-back correlations \cite{Dijets1,Dijets2} or single hadron suppression as a function of the emission angle with the reaction plane \cite{PHENIX-RP}.

Calculations have now reached a high degree of sophistication. Different energy loss formalisms are used together with a 3-d hydrodynamical description of the medium \cite{Hydro3d} in central and noncentral collisions to determine the pathlength dependence of energy loss \cite{HydroJet1,HydroJet2,HydroJet3,HydroJet4}. Some of these models have also been employed successfully to describe the suppression of hard back-to-back hadron correlations \cite{Dihadron1,Dihadron2,Dihadron3}.

The existing formulations of energy loss can roughly be divided into two groups: Some compute the energy loss from a leading parton \cite{Jet2,Jet5,QuenchingWeights} whereas others compute an in-medium fragmentation function by following the evolution of a parton shower \cite{HydroJet2,HBP}. Recently, the Monte Carlo (MC) code JEWEL \cite{JEWEL} has also been developed which simulates the evolution of a parton shower in the medium in a non-analytic way. This model builds on the success of MC shower generators like PYTHIA \cite{PYTHIA,PYSHOW} or HERWIG \cite{HERWIG} for showers in vacuum.

In the present work, we follow an approach which is very similar to the one taken with JEWEL, i.e.\ we modify a MC code for vacuum shower to account for medium effects. However, while JEWEL so far chiefly implements elastic scattering with medium constituents and includes radiative energy loss only in a schematic way, we rather wish to focus on radiative energy loss in the following. This is based on the observation that elastic energy loss has the wrong pathlength dependence to account properly for the suppression of back-to-back correlations \cite{Elastic} and hence cannot be a large contribution to the total energy loss of light quarks. In particular, we assume that partons traversing the medium pick up additional virtuality which induces additional branchings in the shower, thus softening the parton spectrum, but that there is no transfer of longitudinal momentum from the hard parton to the scatterers in the medium.

This work is organized as follows: First, we outline the computation of hadron spectra in the formalism, starting from the hard process. The key ingredient of our model, the medium-modified fragmentation function (MMFF) is described in detail in section \ref{S-MMFF} where we outline the MC simulation of showers in vacuum and present how the algorithm is modified to simulate showers in medium. We present various observables which show the expected modification of the jet by the medium. In section \ref{S-Data} we present a comparison of the suppression calculated using the MMFF in a 3-d hydrodynamical model for the medium evolution \cite{Hydro3d} with the measured nuclear suppression in central AuAu collisions at 200 AGeV and use this result to extract an estimate for the medium transport coefficient $\hat{q}$. We follow with a discussion of the model uncertainties. Finally the limits of the approach in the light of the patterns seen in semi-hard and soft correlations with a hard trigger hadron are discussed.

\section{The hard process}

We aim at a description of the production of high $P_T$ hadrons both in p-p and in Au-Au collisions. The underlying hard process
can be computed in leading order (LO) pQCD. We assume in the following that the actual
hard process is not influenced by the fact that a soft medium is created in Au-Au collisions, but that the subsequent parton shower (which extends to timescales at which a medium is relevant) is modified by the presence of such a medium, whereas hadronization itself takes place sufficiently away from the medium such that it can be assumed to take place as in vacuum. In this section, we describe the computation of the hard process itself.

The production of two hard back to back partons $k,l$ with momentum $p_T$ in a p-p or A-A collision in LO pQCD is described by
 
\begin{equation}
\label{E-2Parton}
\frac{d\sigma^{AB\rightarrow kl +X}}{d p_T^2 dy_1 dy_2} \negthickspace = \sum_{ij} x_1 f_{i/A} 
(x_1, Q^2) x_2 f_{j/B} (x_2,Q^2) \frac{d\hat{\sigma}^{ij\rightarrow kl}}{d\hat{t}}
\end{equation}
 
where $A$ and $B$ stand for the colliding objects (protons or nuclei) and $y_{1(2)}$ is the 
rapidity of parton $k(l)$. The distribution function of a parton type $i$ in $A$ at a momentum 
fraction $x_1$ and a factorization scale $Q \sim p_T$ is $f_{i/A}(x_1, Q^2)$. The distribution 
functions are different for the free protons \cite{CTEQ1,CTEQ2} and protons in nuclei 
\cite{NPDF,EKS98}. The fractional momenta of the colliding partons $i$, $j$ are given by
$ x_{1,2} = \frac{p_T}{\sqrt{s}} \left(\exp[\pm y_1] + \exp[\pm y_2] \right)$.
 
Expressions for the pQCD subprocesses $\frac{d\hat{\sigma}^{ij\rightarrow kl}}{d\hat{t}}(\hat{s}, 
\hat{t},\hat{u})$ as a function of the parton Mandelstam variables $\hat{s}, \hat{t}$ and $\hat{u}$ 
can be found e.g. in \cite{pQCD-Xsec}. Inclusive production of a parton flavour $f$ at rapidity 
$y_f$ is found by integrating over either $y_1$ or $y_2$ and summing over appropriate combinations 
of partons,
 
\begin{widetext}
\begin{equation}
\label{E-1Parton}
\begin{split}
\frac{d\sigma^{AB\rightarrow f+X}}{dp_T^2 dy_f}  = \int d y_2 \sum_{\langle ij\rangle, \langle kl  
\rangle} \frac{1}{1+\delta_{kl}} \frac{1}{1+\delta_{ij}} &\Bigg\{ x_1 f_{i/A}(x_1,Q^2) x_2 
f_{j/B}(x_2,Q^2) \bigg[ \frac{d\sigma^{ij\rightarrow kl}}{d\hat{t}}(\hat{s}, \hat{t},\hat{u})  
\delta_{fk} +
\frac{d\sigma^{ij\rightarrow kl}}{d\hat{t}}(\hat{s}, \hat{u},\hat{t}) \delta_{fl} \bigg]\\
+&x_1 f_{j/A}(x_1,Q^2) x_2 f_{i/B}(x_2,Q^2) \bigg[ \frac{d\sigma^{ij\rightarrow kl}}{d\hat{t}}(\hat{s},  
\hat{u},\hat{t}) \delta_{fk} +
\frac{d\sigma^{ij\rightarrow kl}}{d\hat{t}}(\hat{s},\hat{t}, \hat{u}) \delta_{fl} \bigg] \Bigg\} \\
\end{split}
\end{equation}
\end{widetext}
 
where the summation $\langle ij\rangle, \langle kl \rangle$ runs over pairs $gg, gq, g\overline{q}, 
qq, q\overline{q}, \overline{q}\overline{q}$ and $q$ stands for any of the quark flavours $u,d,s$.

For the production of a hadron $h$ with mass $M$, transverse momentum $P_T$ at rapidity $y$ and 
transverse mass $m_T = \sqrt{M^2 + P_T^2}$ from the parton $f$, let us introduce the fraction $z$ 
of the parton energy carried by the hadron after fragmentation with $z = E_h/E_f$. Assuming 
collinear fragmentation, the hadronic variables can be written in terms of the partonic ones as
 
\begin{equation}
m_T \cosh y = z p_T \cosh y_f \quad \text{and} \quad m_T \sinh y = P_T \sinh y_f.
\end{equation}
 
Thus, the hadronic momentum spectrum arises from the partonic one by folding with the  
distribution $D_{f\rightarrow h}(z, \mu_f^2)$ which describes the average number of fragment hadrons carrying  
a fraction $z$ of the parton momentum at a scale $\mu_f 
\sim P_T$ as
 
\begin{widetext}
\begin{equation}
\label{E-Fragment}
\frac{d\sigma^{AB\rightarrow h+X}}{dP_T^2 dy} = \sum_f \int dp_T^2 dy_f  
\frac{d\sigma^{AB\rightarrow f+X}}{dp_T^2 dy_f} \int_{z_{min}}^1 dz D_{f\rightarrow h}(z, \mu_f^2)  
\delta\left(m_T^2 - M_T^2(p_T, y_f, z)\right) \delta\left(y - Y(p_T, y_f,z)\right)
\end{equation} 
\end{widetext}
 
with
 
\begin{equation}
M_T^2(p_T, y_f, z) = (zp_T)^2 + M^2 \tanh^2 y_f
\end{equation}
 
and 
\begin{equation}
 Y(p_T, y_f, z) = \text{arsinh} \left(\frac{P_T}{m_T} \sinh y_f \right).
\end{equation}
 
The lower cutoff $z_{min} = \frac{2 m_T}{\sqrt{s}} \cosh y$ arises 
from the fact that there is a kinematical limit on the parton momentum; it cannot exceed 
$\sqrt{s}/(2\cosh y_f)$ and thus for given hadron momentum there is a minimal $z$ 
corresponding to fragmentation of a parton with maximal momentum. 

In the following, we describe this key ingredient in our formulation of energy loss, i.e. the computation the fragmentation function both in vacuum and in medium using a pQCD shower evolution code.

\section{The medium-modified fragmentation function}

\label{S-MMFF}

In this section, we describe how the medium-modified fragmentation function is obtained from a computation of an in-medium shower followed by hadronization. Key ingredient for this computation is a pQCD MC shower algorithm. In this work, we employ a modification of the PYTHIA shower algorithm PYSHOW \cite{PYSHOW}. In the absence of any medium effects, our algorithm therefore corresponds directly to the PYTHIA shower. Furthermore, the subsequent hadronization of the shower is assumed to take place outside of the medium, even if the shower itself was medium-modified. It is computed using the Lund string fragmentation scheme \cite{Lund} which is also part of PYTHIA.

\subsection{Shower evolution in vacuum}

We model the evolution from some initial, highly virtual parton to a final state parton shower as a series of branching processes $a \rightarrow b+c$ where $a$ is called the parent parton and $b$ and $c$ are referred to as daughters. In a longer chain of branchings, all partons which have been part of the evolution branch towards some parton $i$ are called the ancestors of $i$ in the following. 

 In QCD, the allowed branching processes are $q \rightarrow qg$, $g \rightarrow gg$ and $g \rightarrow q \overline{q}$.  The kinematics of a branching is described in terms of the virtuality scale $Q^2$ and of the energy fraction $z$, where the energy of daughter $b$ is given by $E_b = z E_a$ and of the daughter $c$ by $E_c = (1-z) E_a$. It is convenient to introduce $t = \ln Q^2/\Lambda_{QCD}$ where $\Lambda_{QCD}$ is the scale parameter of QCD. $t$ takes a role similar to a time in the evolution equations, as it describes the evolution from some high initial virtuality $Q_0$ ($t_0$) to a lower virtuality $Q_m$ ($t_m$) at which the next branching occurs. In terms of the two variables, the differential probability $dP_a$ for a parton $a$ to branch is \cite{DGLAP1,DGLAP2}

\begin{equation}
dP_a = \sum_{b,c} \frac{\alpha_s}{2\pi} P_{a\rightarrow bc}(z) dt dz
\end{equation}

where $\alpha_s$ is the strong coupling and the splitting kernels $P_{a\rightarrow bc}(z)$ read

\begin{eqnarray}
&&P_{q\rightarrow qg}(z) = 4/3 \frac{1+z^2}{1-z} \label{E-qqg}\\
&&P_{g\rightarrow gg}(z) = 3 \frac{(1-z(1-z))^2}{z(1-z)}\label{E-ggg}\\
&&P_{g\rightarrow q\overline{q}}(z) = N_F/2 (z^2 + (1-z)^2)\label{E-gqq}
\end{eqnarray}

where we do not consider electromagnetic branchings involving a photon or a lepton pair and $N_F$ counts the number of active quark flavours for given virtuality. 

At a given value of the scale $t$, the differential probability for a branching to occur is given by the integral over all allowed values of $z$ in the branching kernel as

\begin{equation}
I_{a\rightarrow bc}(t) = \int_{z_-(t)}^{z_+(t)} dz \frac{\alpha_s}{2\pi} P_{a\rightarrow bc}(z).
\end{equation}

The kinematically allowed range of $z$ is given by

\begin{widetext}
\begin{equation}
\label{E-KB}
z_\pm = \frac{1}{2} \left( 1+ \frac{M_b^2 - M_c^2}{M_a^2}\pm \frac{|{\bf p}_a|}{E_a}\frac{\sqrt{(M_a^2-M_b^2-M_c^2)^2 -4M_b^2M_c^2}}{M_a^2} \right)
\end{equation}
\end{widetext}

where $M_i^2 = Q_i^2 + m_i^2$ with $m_i$ the bare quarm mass or zero in the case of a gluon. Note that the gluon emission from a quark Eq.~(\ref{E-qqg}) is singular at $z=1$ and the gluon splitting probability (\ref{E-ggg}) at $z=1$ and $z=0$, hence they lead preferably to soft gluon radiation close to the kinematic cutoff $z_\pm$. Note also that a high virtuality $Q_a$ shrinks the available range of $z$ and hence enforces on average harder radiation.

Given the initial parent virtuality $Q_a^2$ or equivalently $t_a$, the virtuality at which the next branching occurs can be determined with the help of the Sudakov form factor $S_a(t)$, i.e. the probability that no branching occurs between $t_0$ and $t_m$, where

\begin{equation}
S_a(t) = \exp\left[ - \int_{t_0}^{t_m} dt' \sum_{b,c} I_{a \rightarrow bc}(t') \right].
\end{equation}

Thus, the probability density that a branching of $a$ occurs at $t_m$ is given by

\begin{equation}
\label{E-Qsq}
\frac{dP_a}{dt} = \left[\sum_{b,c}I_{a\rightarrow bc}(t)  \right] S_a(t).
\end{equation}

These equations are solved for each branching by the PYSHOW algorithm \cite{PYSHOW} iteratively to generate a shower. For each branching first Eq.~(\ref{E-Qsq}) is solved to determine the scale of the next branching, then Eqs.~(\ref{E-qqg})-(\ref{E-gqq}) are evaluated to determine the type of branching and the value of $z$, if the value of $z$ is outside the kinematic bound given by Eq.~(\ref{E-KB}) then the event is rejected. Given $t_0, t_m$ and $z$, energy-momentum conservation determines the rest of the kinematics except for a radial angle by which the plane spanned by the vectors of the daughter parents can be rotated.

In order to account in a schematic way for higher order interference terms, angular ordering is enforced onto the shower, i.e. opening angles spanned between daughter pairs $b,c$ from a parent $a$ are enforced to decrease according to the condition

\begin{equation}
\label{E-Angular}
\frac{z_b (1-z)b)}{M_b^2} > \frac{1-z_a}{z_a M_a^2}
\end{equation}

After a branching process has been computed, the same algorithm is applied to the two daughter partons treating them as new mothers. The branching is continued down to a scale $Q_{min}$ which is set to 1 GeV in the MC simulation, after which the partons are set on-shell, adjusting transverse momentum to ensure energy-momentum conservation.

\subsection{Shower evolution in medium}

We now embed this shower evolution in a soft medium. It is important to realize that this implies that energy and momentum in this case is in general {\em not} conserved if one includs only one component (shower or medium) in the simulation, as there is flow of energy and momentum through interactions from the shower to the medium and vice versa.

In this particular model, we aim at a description of radiative energy loss, i.e.\ we assume a medium which does not absorb momentum by recoil of its constituents, but rather increases the virtuality of partons propagating through it and thus inducing additional radiation. While this is clearly an idealized assumption of the medium, it allows to study cleanly the effect of induced radiation on a shower, and moreover it is expected to capture the physics of the dominant component of energy loss \cite{Elastic}. Such a medium can be characterize by a transport coefficient $\hat{q}$ which represents the increase in virtuality $\Delta Q^2$ per unit pathlength of a parton traversing the medium. Note that this represents an average transfer, i.e.\ a picture which would be realized in a medium which is characterized by multiple soft scatterings with the hard parton. Alternatively, a model in which hard scatterings which occur probabilistically transfer the virtuality could be implemented. 

While the vacuum shower develops only in momentum space, the interaction with the medium requires modelling in position space as well. In order to make the link, we assume that the formation time of a shower parton with virtuality $Q$ is developed on the timescale $1/Q$, i.e. the lifetime of a virtual parton with virtuality $Q_b$ coming from a parent parton with virtuality $Q_a$ is in the rest frame of the original hard collision (the rest frame of the medium may be different by a flow boost as the medium may not be static) given by

\begin{equation}
\label{E-Lifetime}
\tau_b = \frac{E_b}{Q_b^2} - \frac{E_b}{Q_a^2}.
\end{equation}  

The time $\tau_a^0$ at which a parton $a$ is produced in a branching can be determine by summing the lifetimes of all ancestors. Thus, during its lifetime, the parton virtuality is increased by the amount

\begin{equation}
\label{E-Qgain}
\Delta Q_a^2 = \int_{\tau_a^0}^{\tau_a^0 + \tau_a} d\zeta \hat{q}(\zeta)
\end{equation}

where $\zeta$ is integrated along the spacetime path of the parton through the medium and $\hat{q}(\zeta) = \hat{q}(\tau,r,\phi,\eta_s)$ describes the spacetime variation of the transport coefficient where $\hat{q}$ is specified at each set of coordinates proper time $\tau$, radius $r$, angle $\phi$ and spacetime rapidity $\eta_s$. In the following, we will use a hydrodynamical evolution model of the medium \cite{Hydro3d} for this dependence. If the parton is a gluon, the virtuality transfer from the medium is increased by the ratio of their Casimir color factors, $3/\frac{4}{3} =  2.25$.

If $\Delta Q_a^2 \ll Q_a^2$, holds, i.e. the virtuality picked up from the medium is a correction to the initial parton virtuality, we may add $\Delta Q_a^2$ to the virtuality of parton $a$ before using Eq.~(\ref{E-Qsq}) to determine the kinematics of the next branching. If the condition is not fulfilled, the lifetime is determined by $Q_a^2 + \Delta Q_a^2$ and may be significantly shortened by virtuality picked up from the medium. In this case we iterate Eqs.~(\ref{E-Lifetime}),(\ref{E-Qgain}) to determine a self-consistent pair of $(\tau_a, \Delta Q^2_a)$. This ensures that on the level of averages, the lifetime is treated consistently with the virtuality picked up from the medium.

The need for the self-consistent iteration must be investigated a posteriori when an actual $\hat{q}(\zeta)$ is specified. At this point, we note that it may frequently occur in a constant medium that virtuality transferred from the medium determines the lifetime of a parton, as towards the end of the shower development $Q^2$ may become small, leading to large lifetimes and large $\Delta Q^2$. But the situation is much less severe for an expanding medium in which $\hat{q} \sim 1/\tau^\alpha$ where $\alpha >1$ and $\tau$ is the lifetime of the medium.

We note that in this model, the shower {\em gains} energy from the medium. This is not unexpected, as we have only included energy transfer to the shower, but no interaction which would feed energy from the shower into the medium. In principle, one may argue that sufficiently soft partons of the shower would thermalize and thus become part of the medium, but at the moment we refrain from modelling energy transfer to the medium. Instead, we note that the increase in parton virtuality leads to induced radiation, transferring energy from hard partons to soft radiation and thus to a softening of the momentum distribution of partons inside the shower, along with an increase in momentum $|q_T|$ transverse to the shower initiating parton direction.

\subsection{Properties of the hydrodynamical medium}

After using the Lund string fragmentation scheme \cite{Lund} to hadronize the shower, we can compute the longitudinal momentum distribution of a shower initiated by a parton at fixed energy and hence the fragmentation function into any hadron species at this given partonic momentum scale.

It is clear that the form of the medium-modified fragmentation function (MMFF) is dependent on $\hat{q}(\zeta)$ and that no general form can be given without specifying this quantity. However, we are only interested in very specific forms of $\hat{q}(\zeta)$, i.e. those which occur in the soft medium created in an heavy-ion collision.

\begin{figure}[htb]
\epsfig{file=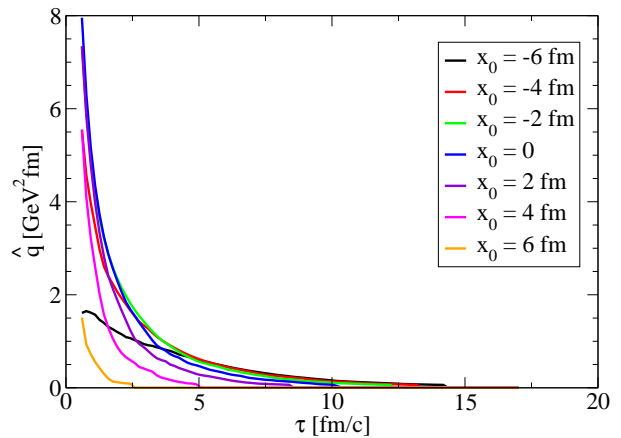, width=8cm}
\caption{\label{F-qhat}Transport coefficient $\hat{q}$ encountered by a quark travelling into $+x$ direction through the hydrodynamical medium in a 200 AGeV collision at impact paramter $b=2.4$ fm for different initial hard vertex position in the transverse $(x,y)$ plane (in all cases $y=0$). Values have been extracted for $K=1.5$ (see Eq.~(\ref{E-qhat}), which represents the best fit to the data.}
\end{figure}

If we assume the relation

\begin{equation}
\label{E-qhat}
\hat{q}(\zeta) = K \cdot 2 \cdot [\epsilon(\zeta)]^{3/4} (\cosh \rho(\zeta) - \sinh \rho(\zeta) \cos\psi)
\end{equation}

between $\hat{q}$ and the medium energy density $\epsilon$, the local flow rapidity $\rho$ with angle $\psi$ between flow and parton trajectory \cite{Flow1,Flow2}, we find that the vast majority of paths leads to a $\hat{q}(\zeta)$ which can be described by (see Fig.~\ref{F-qhat} for examples)

\begin{equation}
\hat{q}(\zeta) = \frac{a}{(b+\tau/(1 fm/c))^c}.
\end{equation}

The exception are paths close to the surface for which the parton travels {\em inward}. However, such events are not only suppressed by the overlap function which is small at the medium surface but also very suppressed due to the strong medium effect for such long paths. Thus, the medium which will be probed by a parton emerging from a particular vertex can be characterized by the four parameters $(a,b,c,\tau_E)$ where $\tau_E$ is the time at which the parton emerges from the medium. In practice, the range of parameters can be sufficiently narrowed down to 1  $<b<$ 2, $\tau_E < \tau_F$ (with $\tau_F$ the freeze-out time of the medium) and $2 < c < 4$.

We thus investigate in the following three different scenarios (approximately representing a parton travelling into $+x$ direction originating from $x=4$ fm (A), $x=0$ (B) and $x=-4$ fm (C), $y=0$ in all cases in the transverse $(x,y)$ plane at midrapidity. These trajectories are characterized by the parameters $(b=1.5, c=3.3, \tau_E= 5.8$ fm/c$)$ (A),  $(b=1.5, c=2.2, \tau_E= 10$ fm/c$)$ (B) and $(b=1.5, c=2.2, \tau_E= 15$ fm/c$)$ (C) and are quite typical for partons close to the surface (A), emerging from the central region (B) or traversing the whole medium (C).

\subsection{The medium-modified fragmentation function}

\label{S-scaling}

\begin{figure}[htb]
\epsfig{file=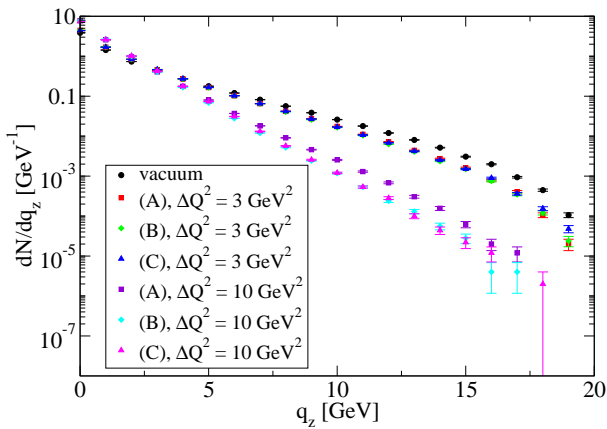, width=8cm}
\caption{\label{F-MMFF}Longitudinal momentum distribution of hadrons in a jet originating from a 20 GeV quark in vacuum and in medium. The medium results are given for two different values of line-integrated virtuality $\Delta Q^2$ for scenarios (A,B,C) describing three different characteristic paths through the medium (see text for details).}
\end{figure}

In Fig.~\ref{F-MMFF}, we present the longitudinal momentum $q_z$ distribution of hadrons in a jet originating from a 20 GeV quark, both in vacuum and medium-modified according to the three scenarios outlined above. Instead of specifying the parameter $a$, we characterize the scenarios instead by the total amount of virtuality

\begin{equation} 
\Delta Q^2_{tot} = \int_{\tau_0}^{\tau_E} d \zeta \hat{q}(\zeta) 
\end{equation}

which would be transferred to a single parton crossing the medium (note that since branchings occur, multiple partons travel through the medium in a shower, hence the virtuality transfer to the shower as a whole is larger than $\Delta Q^2_{tot}$, but can only be computed on an event-by-event basis). 

In general, the medium modification induces the expected changes to the shower, i.e. the high $q_z$ tail is reduced by the medium, whereas induced radiation creates additional multiplicity at low $q_z$. This transfer of energy from hard partons to soft radiation increases with the strength of the medium transport coefficient.

One notes that for small $\Delta Q^2_{tot}$, there is good scaling between all scenarios, and only for larger values of $\Delta Q^2_{tot}$ the precise pattern at which time the virtuality is transferred to the shower starts to be significant and longer paths lead to more suppression of the high $q_T$ tail. However, since partons close to the edge of the medium propagating outward traverse only low density matter, they probe a region in which the scaling holds. Thus, scenario (B) can effectively be assumed to be a fair approximation for all paths if $\Delta Q^2_{tot}$ is used to characterize the path. This observation simplifies the subsequent averaging of paths through the hydrodynamical medium considerably, as it reduces the problem to determining $\Delta Q^2_{tot}$ on line integrals through the medium.

\subsection{The Hump-backed plateau}

In order to focus more on the hadron production at low $p_z$, we introduce the variable $\xi = \ln(1/x)$ where $x = p/E_{jet}$ is the fraction of the jet momentum carried by a particular hadron and $E_{jet}$ is the total energy of the jet. The inclusive distribution $dN/d\xi$, the so-called Hump-backed plateau, is an important feature of QCD radiation \cite{Muller,Dokshitzer} and is in vacuum dominated by color coherence physics.

\begin{figure}[htb]
\epsfig{file=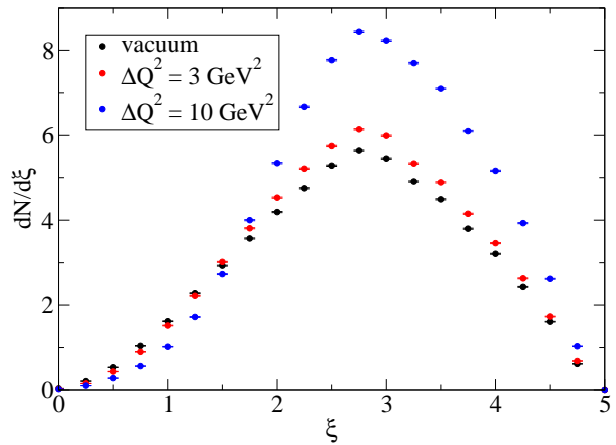,width=8cm}
\caption{\label{F-HBP}The Hump-backed plateau distribution $dN/d\xi$ for a jet originating from a 20 GeV quark shown both in vacuum and medium modified at two different values of line-integrated virtuality $\Delta Q^2$. }
\end{figure}

Fig.~\ref{F-HBP} shows the calculated distribution in vacuum and its distortion by the medium effect. Qualitatively, the results are very similar to those observed in \cite{HBP} and \cite{JEWEL} for schematic radiative energy loss, i.e.\ a depletion of the distribution at low $\xi$ and a strong enhancement of the hump.

\subsection{Transverse jet momentum spectra and angular distribution}

\begin{figure}[htb]
\epsfig{file=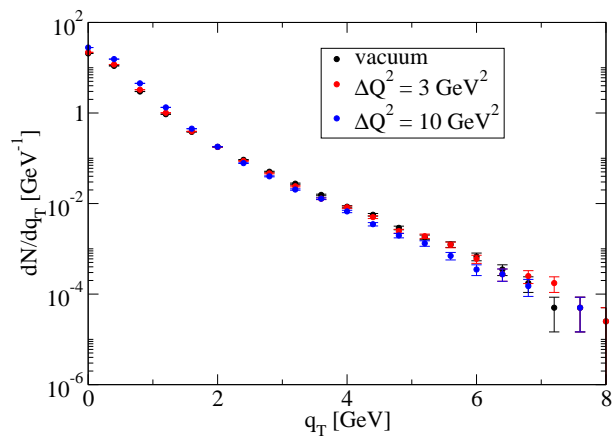, width=8cm}
\caption{\label{F-pT}Momentum distribution of hadrons in the jet transverse to the direction of the shower initiating parton for a 20 GeV quark, both in vacuum and in medium for two different values of the line-integrated virtualtiy $\Delta Q^2$. }
\end{figure}

Since the parton shower picks up virtuality from the medium, we may expect some broadening of the transverse distribution of hadrons in the jet due to the medium effect. We show the transverse momentum $q_T$ distribution (where transverse in this section denotes the direction transverse to the hard parton initiating the shower, not as in a description of the whole p-p or Au-Au collision, the direction transverse to the beam axis) of hadrons in the shower in Fig.~\ref{F-pT}.

Although the effects on the transverse momentum spectrum are not large, there is a clear trend to a rise of multiplicity at low $q_T$ and some depletion in the high $q_T$ tail through the medium effect. One may also note that the integral $\int dq_T q_T dN/dq_T$ increases by $\sim 30$\% from vacuum to the $\Delta Q^2_{tot}=10$ GeV$^2$ medium modified jet.

\begin{figure}[htb]
\epsfig{file=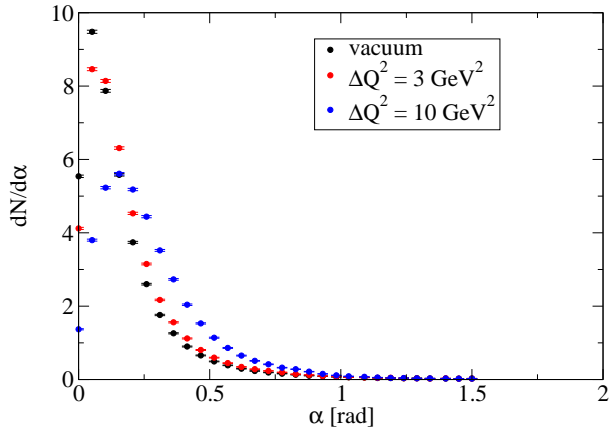, width=8cm}
\caption{\label{F-ang}Angular distribution of hadrons with respect to the jet axis for a shower initiated by a 20 GeV quark, both in vacuum and in medium for two different values of the line-integrated virtualtiy $\Delta Q^2$. }
\end{figure}

The effect of the medium on the transverse structure of the shower is more clearly seen when looking at the angular distribution of hadrons in the jet. The distribution $dN/d\alpha$ where $\alpha$ is the angle between hadron and jet axis is shown in Fig.~\ref{F-ang} where a cut in momentum of 2 GeV has been applied to focus on hadrons which would appear above the soft background of a heavy-ion collision.

The angular broadening of the distribution by the medium is clearly visible (note that the dip at $\alpha = 0$ is caused by the Jacobian in a polar coordinate system and is not an indication of a double-hump angular structure).

\section{Computation of $R_{AA}$}

In this section, we show how to compute the nuclear suppression factor $R_{AA}$ using the medium-modified fragmentation function in the hydrodynamically evolving medium.
The nuclear suppression factor is defined as
\begin{equation}
R_{AA}(P_T,y) = \frac{dN^h_{AA}/dP_Tdy}{T_{AA}(0) d\sigma^{pp}/dP_Tdy}
\end{equation}

where $T_{AA}({\bf b})$ is the standard nuclear overlap function. We can compute it by forming the ratio 

\begin{equation}
\label{E-R_AA}
R_{AA}(P_T,y) = \frac{d\tilde\sigma_{medium}^{AA\rightarrow h+X}}{dP_T^2  
dy}/\frac{d\sigma^{pp\rightarrow h+X}}{dP_T^2 dy}
\end{equation}

where ${d\sigma^{pp\rightarrow h+X}}/{dP_T^2 dy}$ follows from Eq.~(\ref{E-Fragment}) when $D_{f\rightarrow h}(z, \mu_f^2)$ is set to be the vacuum fragmentation function whereas ${d\tilde\sigma_{medium}^{AA\rightarrow h+X}}/{dP_T^2 dy}$ is computed from the same equation with $D_{f\rightarrow h}(z, \mu_f^2)$ replace by the suitably averaged MMFF $\langle D_{MM}(z,\mu_f^2)\rangle_{T_{AA}}$. This averaging has to be done over all possible paths of partons through the medium.  We make the simplifying assumption that instead of computing the detailed paths of all partons in a shower through the medium, we only sample the medium along the eikonal path of the shower originator and consider this the medium seen by all partons in the shower. In essence, this neglects the transverse spread of paths in a shower. 

The probability density $P(x_0, y_0)$ for finding a hard vertex at the 
transverse position ${\bf r_0} = (x_0,y_0)$ and impact 
parameter ${\bf b}$ is given by the product of the nuclear profile functions as
\begin{equation}
\label{E-Profile}
P(x_0,y_0) = \frac{T_{A}({\bf r_0 + b/2}) T_A(\bf r_0 - b/2)}{T_{AA}({\bf b})},
\end{equation}
where the thickness function is given in terms of Woods-Saxon the nuclear density
$\rho_{A}({\bf r},z)$ as $T_{A}({\bf r})=\int dz \rho_{A}({\bf r},z)$. The MMFF must then be averaged over this quantity and all possible directions $\phi$ partons could travel from a vertex as

\begin{widetext}
\begin{equation}
\label{E-P_TAA}
\langle D_{MM}(z,\mu^2) \rangle_{T_{AA}} \negthickspace = \negthickspace \frac{1}{2\pi} \int_0^{2\pi}  
\negthickspace \negthickspace \negthickspace d\phi 
\int_{-\infty}^{\infty} \negthickspace \negthickspace \negthickspace \negthickspace dx_0 
\int_{-\infty}^{\infty} \negthickspace \negthickspace \negthickspace \negthickspace dy_0 P(x_0,y_0)  
D_{MM}(z,\mu^2,\zeta).
\end{equation}
\end{widetext}

Using the approximate scaling relation described in section \ref{S-scaling}, the medium modified fragmentation function $D_{MM}(z, \mu^2,\zeta)$ for a path $\zeta$ can be found by computing the line intergal

\begin{equation}
\Delta Q^2_{tot} = \int_0^\infty \negthickspace d \zeta \hat{q}(\zeta)
\end{equation}

through the hydrodynamical medium where $\hat{q}(\zeta)$ is given by Eq.~(\ref{E-qhat}). The MC shower code is then used to compute $D_{MM}(z, \mu^2,\zeta)$ for each value of $\Delta Q^2_{tot}$.

Note that this point that there is in principle a conceptual difference between using a fragmentation function extracted from a shower simulation (be it in medium or vacuum) and a phenomenological fragmentation function such as the KKP \cite{KKP} or AKK \cite{AKK} set in Eq.~(\ref{E-Fragment}). This has to do with the fact that in the the usual fragmentation functions $\mu_f$ represents the relevant {\em hadronic} momentum scale. On the other hand, one can extract fragmentation functions from a shower evolution only for a given {\em partonic} momentum scale. Thus, the scale evolution of the fragmentation function must be accounted for in a different way. At present, we use a MMFF for a fixed partonic scale of 20 GeV, hence the description of high $P_T$ hadron momentum spectra with the shower generator extracted fragmentation function is not as good as using the phenomenological functions. However, here we are chiefly interested in the ratio of modified over unmodified spectra, and in this ratio the uncertainties due to the scale evolution largely cancel out. 

\section{Comparison with data}

\label{S-Data}

\begin{figure}[htb]
\epsfig{file=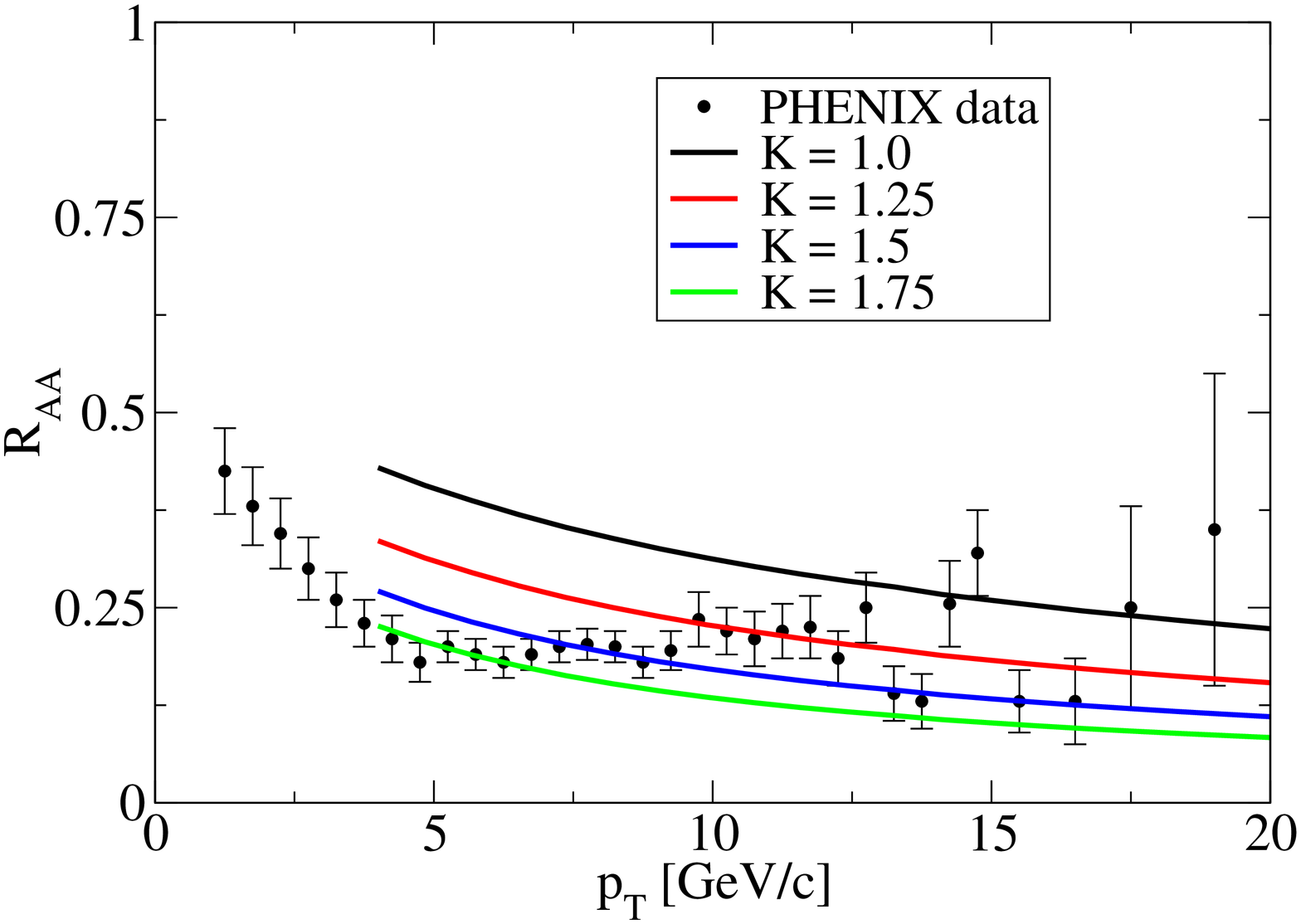,width=8cm}
\caption{\label{F-R_AA}The nuclear suppression factor $R_{AA}$ for central AuAu collisions at 200 AGeV computed for different values of the parameter $K$ specifying the transport coefficient of the medium in comparison with the PHENIX data \cite{PHENIX_R_AA}. }
\end{figure}

At this point, our the model is completely determined except for the parameter $K$ in Eq.~(\ref{E-qhat}) which links the energy density in the medium with the transport coefficient. We adjust $K$ such that the best fit to the data is achieved. In Fig.~\ref{F-R_AA} we show $R_{AA}$ for different values of $K$. 

The data seem to favour $K=1.5$. This implies that the medium exhibits a 50\% stronger effect than expected in an ideal QGP \cite{Baier}. $\hat{q}_0$, i.e. the highest transport coefficient reached in the evolution in the medium center at thermalization time is found to be 7.8 GeV$^2$/fm. Note that this is well below the values of $K$ (and $\hat{q}$) extracted from computations which consider energy loss from the leading parton in the ASW \cite{QuenchingWeights} formulation of energy loss, cf.  where $K$ values between 2 and 5 are found \cite{HydroJet1,Dijets2,JetFlow}.

However, it is evident that the overall trend of the data is not well described by the model. While the data indicate a small rise of $R_{AA}$ from $p_T=5$ GeV and above, the MMFF model finds a drop. In particular, the region between 10 and 14 GeV is not well described. This is quite unlike models based on energy loss of the leading parton where a rise of $R_{AA}$ with $p_T$ is found, cf. e.g. \cite{HydroJet1,Dijets2}. On the other hand, a similar drop of $R_{AA}$ has been found in the schematical shower evolution model in \cite{HBP}. 

It seems clear that there is a direct link between the characteristic distortion of the Hump-backed plateau at larger $\xi$ by medium effects and the rise of $R_{AA}$ towards lower $p_T$ --- in both cases, the effects of additional parton production at lower momenta by medium induced radiation become visible. It seems at this point that this is a generic, unavoidable feature of the model formulation. Nevertheless, let us revisit the assumptions of the model and discuss their validity.

\section{Discussion of model assumptions}

While a detailed discussion of all model uncertainties is beyond the scope of this paper and will be postponed to a future publication, we will at this point provide a list of possible uncertainties and estimate their effect if possible.

\begin{figure}[htb]
\epsfig{file=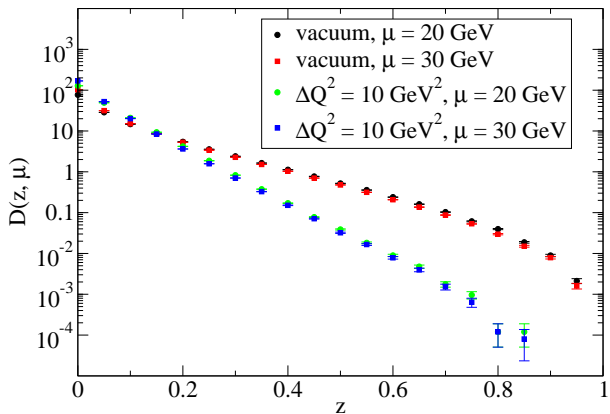,width=8cm}
\caption{\label{F-Scale}Dependence of the fragmentation function, both in vacuum and in medium for a line-integrated virtuality $\Delta Q^2 = 10$ GeV$^2$ for two different values of the partonic momentum scale $\mu$.} 
\end{figure}

First, there are uncertainties in the calculation of the hard process. We do this in LO pQCD, and one may wonder about NLO corrections. However, in \cite{LOpQCD} is has been shown that LO pQCD supplemented with a $K$ factor to account for higher order effects is in fact a good description of hard hadron production in p-p collisions. Another approximation we have made is to neglect the scale dependence of the fragmentation function and instead to compute the fragmentation function at a single partonic scale of 20 GeV. While there is no conceptual problem with a MC computation of the dependence of the vacuum or medium-modified fragmentation on $(z, \mu)$ and in the latter case also $\Delta Q^2_{tot}$, mapping out a dependence on three parameters is very demanding in terms of MC simulation time and will be done in a future investigation. As shown in Fig.~\ref{F-Scale} the effects are not substantial.

A second class of uncertainties has to do with the way the shower algorithm is implemented and affects both vacuum and medium results. In essence, these are encoded in the control parameters of the PYTHIA shower algorithm. For example, the scale $Q^2_{min}$ down to which partons are evolved is a parameter, and the results show some dependence to it. Other uncertainties have to do with what scale is used to determine the running of $\alpha_s$ in the shower. In general, these effects are known to be small.

\begin{figure}[htb]
\epsfig{file=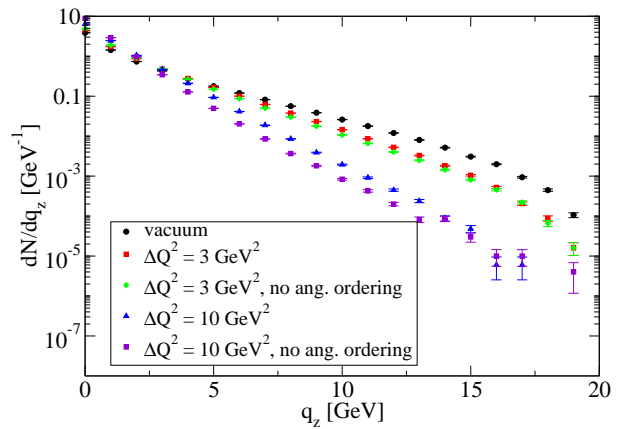,width=8cm}
\caption{\label{F-AngOrder}Longitudinal momentum distribution of hadrons in a jet originating from a 20 GeV quark in vacuum and for two different values of the line-integrated virtuality $\Delta Q^2$. Shown is the effect of retaining angular ordering in the branching (as in vacuum) or dropping it under the assumption that frequent medium interactions destroy coherence. } 
\end{figure}

A third class of uncertainties has to do with the specific implementation of medium effects. Chiefly, one may wonder about the effect of replacing a probabilistic formation time with its average value in Eq.~(\ref{E-Lifetime}) or about replacing a probabilistic transfer of virtuality associated with scattering processes with an average growth of virtuality as realized in Eq.~(\ref{E-Qgain}). The latter assumption has some impact on the treatment of QCD coherence effects, as the angular ordering Eq.~(\ref{E-Angular}) would be destroyed by a scattering process. Thus, a detailed formulation in terms of individual scatterings with medium constituents would include the information when angular ordering is effective and when not. This information is lost when computing based on an average virtuality transfer. On the other hand, Landau-Pomeranchuk-Migdal (LPM) interference \cite{LPM} would lead to a suppression of radiation as compared to the vacuum situation due to frequent interactions with the medium. While answering the question about the effect of replacing a probabilistic formulation with averages requires a restructured algorithm which is beyond the scope of this paper, the effect of a detailed treatment of the angular ordering can be estimated comparatively easy by computing in the two limits of retaining or dropping angular ordering completely in the medium. As evident from Fig.~\ref{F-AngOrder}, the effects are in practice relatively small and moreover do not alter the shape of the distribution significantly, hence they can largely be absorbed into a rescaling of $\Delta Q^2_{tot}$ or alternatively $K$.

One may also keep in mind that using the Lund string fragmentation mechanism to hadronize in a heavy-ion environment is not straightforward, especially at low $z$ and for heavy hadrons. If one estimates the hadronization time for a hadron with mass $m_h$ and energy $E_h$ as

\begin{equation}
\tau \sim E_h/m_h^2
\end{equation} 

then a 10 GeV pion will be formed several tens of fm away from the medium, but a 5 GeV proton has a formation time of about 1.2 fm, clearly not enough to assume the process happens out of the medium or even out of the partonic region. Thus, there is some reason to suspect that the treatment of low $z$ hadrons will miss out part of the relevant physics.

Finally, one may question the validity of the scaling assumption made in section \ref{S-scaling} which allows in essence to reduce the information of $\hat{q}(\zeta)$ specified on the whole path to a simple line integral. While the scaling is not exact, a detailed comparison of Eq.~(\ref{E-Qgain}) for some selected explicit forms of paths through the hydro medium with the scaled version shows that differences in the resulting MMFF are at most on the level of 20\% and moreover can again largely be absorbed into a rescaling of $\Delta Q^2_{tot}$ as the shape of the MMFF is not changed. 

\section{Conclusions}

There is evidence that model uncertainties which can be estimated maninly affect the precise determination of $K$, but not the more striking fact that model does not capture the rising trend of $R_{AA}$ with $p_T$ apparent in the data. While there is still the possibility that one of the effects which have been neglected in modelling so far may correct this disagreement, one may note that this does not seem likely as there is within the model a rather generic physics mechanism which produces this particular shape, i.e. the energy transport from high $z$ partons into radiated low $z$ multiplicity, and it is this additional multiplicity which leads to the rise of $R_{AA}$ for low momenta. Supporting evidence for this is found in the fact that the calculation in \cite{HBP} which is based on a completely different algorithm but models the same physics mechanism results nevertheless in a very similar shape of $R_{AA}$ when folded with the pQCD parton spectrum.

At this point, we note that there's a second failure of the model, also connected with subleading hadrons in the shower. The angular distribution of hadrons shown in Fig.~\ref{F-ang} shows, even for a jet emerging from the center of the medium, only moderate broadening, but no sign of an energy momentum flow to large angles or the formation of a dip at zero angles as would be required to account for experimental data on two and three particle correlations \cite{PHENIX_Mach,PHENIX_Mach2,STAR_Mach,STAR_3p}. In contrast, models which assume that the energy radiated from the leading parton is not developing as a shower but rather excites a hydrodynamical shockwave in the medium \cite{MachShuryak,Mach1,Mach2,Mach3} can account for the observed structures in the data.

On the other hand, radiative energy loss considered for the leading parton only is able to account well for the data even in the nontrivial case of back-to-back correlations \cite{HydroJet1,Dihadron1,Dihadron2}. Furthermore, conceptually branchings at large $Q^2$ should take place regardless of the presence or absence of a medium, as they on average develop before a thermalized medium can be assumed to exist.

These observations lead to the idea that while especially the initial development of a high $p_T$ process is described by the shower evolution, at some lower scale a pQCD shower ceases to be an adequate description for soft radiated partons, but instead some form of a coupling to hydrodynamical modes of the medium (cf. e.g.\cite{Source}) takes place. The interesting question (especially in the light of the kinematic range of hard probes accessible at LHC) is what part of the shower (starting from the leading parton) is still correctly described by the in medium shower evolution equations and under what conditions this evolutions should be replaced by different physics. The answer to these questions at RHIC energies may be found in the application of the formalism presented here to hard multiparticle correlations in which a systematic access to the high $P_T$ part of the shower is accessible. This, however, will be the subject of a future investigation.

\begin{acknowledgments}
I'd like to thank Kari J. Eskola for valuable discussions on the problem. This work was financially supported by the Academy of Finland, Project 115262. 
\end{acknowledgments}

\end{document}